\documentclass[12pt,letterpaper]{article}

\usepackage{amsfonts}
\usepackage{verbatim}
\usepackage[english]{babel}
\usepackage{natbib}
\usepackage{latexsym}
\usepackage{url}
\usepackage{graphicx}
\usepackage{amssymb}
\usepackage{amsmath}
\usepackage{ifthen}

\bibpunct{(}{)}{;}{a}{}{,}  

\begin{document}

\begin{flushright}
Version dated: \today
\end{flushright}
\bigskip

\bigskip
\medskip
\begin{center}

\noindent{\Large \bf Statistical Inference of Allopolyploid Species Networks in the Presence of Incomplete Lineage Sorting}
\bigskip



\noindent {\normalsize \sc Graham Jones$^{1,2}$, Serik Sagitov$^2$, and Bengt Oxelman$^1$}\\
\noindent {\small \it $^1$Department of Biological and Environmental Sciences, University of Gothenburg, Gothenburg, Sweden; $^2$Mathematical Sciences, Chalmers University of Technology and the University of Gothenburg, Gothenburg, Sweden}\\
\end{center}
\medskip
\noindent{\bf Corresponding author:} Graham Jones, 21e Balnakeil, Durness, Lairg,
Sutherland IV27 4PT, UK; E-mail: art@gjones.name.\\

\bigskip
\noindent{{\textbf{Note:} submitted to \textit{Systematic Biology}.}
\bigskip

\begin{center}
\textit{Abstract} 
\end{center}

Polyploidy is an important speciation mechanism, particularly in land plants. Allopolyploid species are formed after hybridization between otherwise intersterile parental species. Recent theoretical progress has led to successful implementation of species tree models that take population genetic parameters into account. However, these models have not included allopolyploid hybridization and the special problems imposed when species trees of allopolyploids are inferred. Here, two new models for the statistical inference of the evolutionary history of allopolyploids are evaluated using simulations and demonstrated on two empirical data sets. It is assumed that there has been a single hybridization event between two diploid species resulting in a genomic allotetraploid. The evolutionary history can be represented as a network or as a multiply labeled tree, in which some pairs of tips are labeled with the same species. In one of the models (AlloppMUL), the multiply labeled tree is inferred directly. This is the simplest model and the most widely applicable, since fewer assumptions are made. The second model (AlloppNET) incorporates the hybridization event explicitly which means that fewer parameters need to be estimated. Both models are implemented in the BEAST framework. Simulations show that both models are useful and that AlloppNET is more accurate if the assumptions it is based on are valid. The models are demonstrated on previously analyzed data from the genus \textit{Pachycladon} (Brassicaceae) and from the genus \textit{Silene} (Caryophyllaceae).\\
\noindent (Keywords: Allopolyploid, hybridization, Bayesian, phylogenetics, network)\\

\bigskip

\section{Introduction}

Polyploidy is an important mechanism for the emergence of new species, which is particularly prominent in plants \citep{Wendel:2005, CuiEtal:2006, WoodEtal:2009}. Allopolyploids are produced by hybridization between two species, and are considerably more common than autopolyploids, which are formed within species \citep{Tate:2005}. Hybridization presents a challenge to phylogenetic analysis since the usual tree is replaced by a network. In addition, it becomes difficult to assign genome identities to allele copies. This imposes a problem for the inference of species trees in a relevant multi-species coalescent framework \citep{YR:2003} even if the hybridization event is ignored. This paper explores the feasibility of making statistical inferences about the evolutionary history of allopolyploids using simulations of some simple scenarios and two novel models implemented in the BEAST software \citep{BEAST}.

The main restrictions made here are that there has been a single hybridization event between two diploids, and that the resulting hybrid is a \textit{genomic allopolyploid}, in which the two diploid genomes (from the two parental diploid species) do not recombine with one another at meiosis because the chromosomes in the two parental species were too diverged by the time the hybrid formed. This leaves two major problems to deal with in the phylogenetic analysis. Firstly, when the DNA from organisms is sequenced, it is not possible to assign sequences to their parental diploid species. Thus, although the sequences can be seen as the result of the evolution of diploid genomes, there is an ambiguity in the labeling of the sequences which is not normally present. Secondly, the issue of incomplete lineage sorting cannot be ignored. 

Figure~1 shows three ways of viewing the same evolutionary events. The three columns show three main scenarios labeled $A$, $B$, and $C$ in which the allotetraploid could have arisen. In all scenarios, a speciation at the root produces two diploids $a$ and $b$, and at some point later, a hybridization occurs between $a$ and $b$ or one or two of their extinct relatives. After hybridization, the tetraploid speciates to produce two species $y$ and $z$. The sections from top to bottom show three different ways of viewing each of these scenarios. In the top section, hybridization and extinction events are explicitly represented. Note that more than one sequence of evolutionary events (speciation, extinction, and hybridization) can correspond to the same representations in the other two views. In the second section, the networks are represented as a collection of homoploid `trees with legs' in which trees of higher ploidy are connected by their legs to those with lower ploidy. The bottom section shows the multiply-labeled tree (MUL-tree) view: there is a binary tree, but some of the tips have the same labels since they correspond to the same species.

\begin{figure}[!hbtp]
\resizebox{1.00\hsize}{!}{\includegraphics*{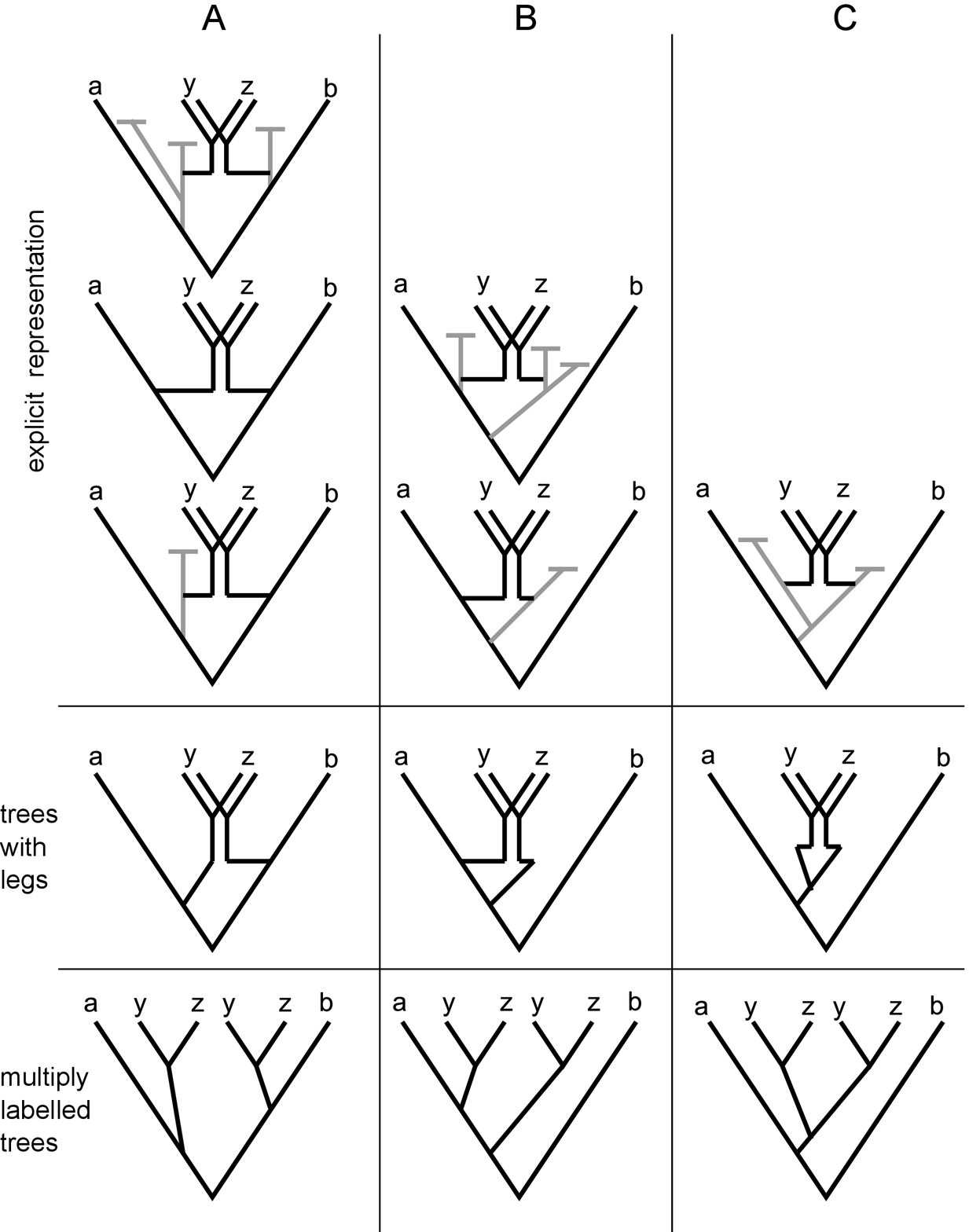}}
\caption{\textsf{The three columns show three main scenarios A,B,C. In each one there are two diploids a and b, and two tetraploids y and z. See main text for more details}.}
\end{figure}

The first view contains the most information, but in general it is difficult or impossible to infer the extra details represented in it. The methods of this paper cannot be used to distinguish whether one or two extinct species hybridized to produce the allotetraploid. It might be possible in scenario $A$ to infer whether there was a direct hybridization between the two extant diploids, or whether extinct (or unsampled) species were involved, using estimates of node heights. It might also be possible to estimate the hybridization time from population sizes, if one assumes that the allotetraploid species arose from a single individual. However it is difficult to obtain good estimates of ages and population sizes. In this paper we focus on inferring the level of detail in the `trees with legs' or MUL-tree views.

The data available for inferring the species history consists of molecular sequences sampled from individuals belonging to species. One approach, pursued in \cite{HuberMoulton:2006}, \cite{HuberOxelman:2006}, and \cite{LottEtal:2009}, is to estimate the gene trees first, and then search for the MUL-tree that best accommodates them. In contrast, the approach here is the typically Bayesian one of `co-estimating everything'. The network node times and topology, the assignment of sequences obtained from tetraploid individuals to parental diploid species, and the node times and topologies of all the gene trees are all allowed to vary, and an MCMC algorithm is used to sample the posterior distribution. The approach is similar to that of *BEAST as described in \cite{HeledDrummond:2010} but the sequence assignment ambiguity is new.

\section{The models}

It is assumed that at some point in the past a diploid species speciated to form two diploid species which both have survived to the present. The initial speciation forms the root of the network or MUL-tree. At some later time a hybridization took place between these two diploid species or their extinct relatives, forming an allotetraploid in which the two parental diploid genomes continue to evolve without recombining with one another. After the hybridization, further speciation of the allotetraploid may have taken place.

Two models, denoted as AlloppMUL and AlloppNET, are considered. They are both based on the multi-species coalescent \citep{YR:2003}. It is assumed that there is free recombination between genes, but no recombination within genes. While the assumption of no recombination within genes may well be unrealistic, simulations conducted in \cite{LanierKnowles:2012} suggest that this violation of the model does not pose a major problem. Also as in \cite{HeledDrummond:2010}, the term species ``is not necessarily the same as a taxonomic rank, but designates any group of individuals that after some `divergence' time, have no history of breeding with individuals outside that group.''.

In the usual formulation of the multi-species coalescent there is a species tree and a number of gene trees, and it is assumed that individuals can be assigned unambiguously to tips in the species tree, and that molecular sequences can also be assigned unambiguously to tips in the species tree. If the first assumption is relaxed, so that the assignment of individuals to species must be estimated, the result is a model like that of \cite{YR:2010} which is used to delimit species. In the models considered here, the second assumption is relaxed, to cater for the ambiguity in assigning multiple sequences from the same individual to tips in the multiply labeled species tree. 

In the first model (AlloppMUL) the multiply labeled tree is inferred directly. The topology, the node times, and the population sizes along the branches are allowed to vary freely, as if the diploid genomes within the allotetraploid(s) belonged to different species. This approach therefore throws away some information implicit in the assumptions. The two main advantages of this approach are that it may be more appropriate where the assumptions are dubious, especially when the number of hybridization events is not known, and the simplicity of implementation (since it is a relatively straightforward generalization of the implementation of the multi-species coalescent model in *BEAST).

The second model (AlloppNET) is more faithful to evolutionary events. The hybridization is modeled explicitly as a node in a network, and from that time, the diploid genomes within the allotetraploid(s) must share population sizes and speciation events. Since this uses more information it is expected to be more accurate. The network can be converted into a MUL-tree for calculations and for program output. The key point is that since the MUL-tree is derived from the network, the appropriate constraints on the topology, node times, and populations are enforced onto the MUL-tree.

\subsection{The AlloppMUL model}

The posterior probability for the AlloppMUL model is given by

\begin{eqnarray}\label{eq:AlloppMUL}
\Pr(M, \theta, \tau, \alpha, \gamma | d) & \propto & \Pr(M|\lambda)\Pr(\lambda) \times \nonumber\\
                              &         & \Pr(\theta|\eta)\Pr(\eta) \times \nonumber\\
                              &         & \Pr(\gamma) \times \nonumber\\
                              &         & \prod_{i=1}^G \Pr(\tau_i|M, \theta, \gamma_i) \times \nonumber\\
                              &         & \prod_{i=1}^G \Pr(d_i|\tau_i, \alpha_i).
\end{eqnarray}
Here the multiply-labelled species tree is denoted by $M$, and the parameter(s) for the topology and node times in the prior for $M$ are denoted by $\lambda$. The population size parameters are denoted by the vector $\theta$. The parameter $\eta$ is a scaling factor for the population sizes, appearing in a hyperprior for $\theta$. The number of gene trees is denoted by $G$. The topology and set of node times for the $i$th gene tree is denoted by $\tau_i$ ($1 \leq i \leq G$). All the other parameters belonging to the $i$th gene tree are denoted by $\alpha_i$; these are parameters for site rate heterogeneity, substitution model, branch rate model, and root model. Thus $(\tau_i,\alpha_i)$ gives all the parameters for the $i$th gene tree. The permutations of sequences within polyploid individuals for the $i$th gene is denoted by $\gamma_i$. This parameter is the main addition to the usual formula for the multispecies coalescent. We only deal with tetraploids here, so $\gamma_i$ consists of transpositions (`flips') of two sequences. The sequence data for the $i$th gene is denoted by $d_i$. We set $\tau = (\tau_1,...\tau_G)$, and similarly for $\alpha, \gamma, d$. The five terms in this expression will now be described in detail.

\begin{itemize}
\item{The species tree prior $\Pr(M|\lambda)\Pr(\lambda)$ is the probability of $M$ before seeing any molecular data. Little is known about what an appropriate prior should be. For the analyses in this paper, $M$ is regarded as an ordinary tree, and a Yule prior is used for this. The single parameter $\lambda$ represents the birth rate.}

\item{The population prior $\Pr(\theta|\eta)\Pr(\eta)$ is for the population size parameters $\theta$. There is one value at each tip, and one at the root-ward end of each branch in the MUL-tree. In the analyses done in this paper, the priors for $\theta$ used were similar to those typically used by *BEAST. An independent gamma distribution is assumed for each population size. The shape parameter is 4 for the populations at the tips, and 2 for the rest. If it is assumed that the total population just before and just after a speciation is the same, then at tip-ward end of an internal branch, the population is the sum of the two gamma distributions with shape parameter 2, and is thus a gamma distribution with shape parameter 4, like the tips. The scale parameter for all these gamma distributions is the hyperparameter $\eta$. The populations are assumed to vary linearly along edges in the network, between the nodes where the population parameters occur. The hyperprior for $\eta$ is described later.}

\item{The permutation prior $\Pr(\gamma)$ is the prior probability of sequence assignments. This is assumed to be uniform here, and thus could be ommitted without affecting the inference.}

\item{The term $\Pr(\tau_i|M, \theta, \gamma_i)$ is the probability of $\tau_i$, when permuted by $\gamma_i$, fitting into the species tree $M$ with populations determined by $\theta$. The value of $\gamma_i$ determines how the sequences for the $i$th gene are assigned to tips in the multiply labeled tree $M$. Note that this probability does not depend on $\alpha_i$. Apart from this extra complexity due to the permutations, the value of $\Pr(\tau_i|M, \theta, \gamma_i)$ is given by the multispecies coalescent, as in \cite{HeledDrummond:2010}.}

\item{The gene tree likelihood $\Pr(d_i|\tau_i, \alpha_i) = \Pr(d_i|\tau_i,\alpha_i)$ is the usual `Felsenstein likelihood' of the data for the $i$th gene given the $i$th gene tree. The parameter $\alpha_i$ is described in more detail in subsection `Other parts of the models' below. It may be helpful to think about the gene tree likelihood and the previous term in another way. One can think of the $\gamma_i$ as permuting the sequence data $d_i$, that is swapping pairs of rows in a data matrix for the $i$th gene, where the pairs have been sequenced from the same individual. Then for each $i$, the product $\Pr(d_i|\tau_i, \alpha_i) \times \Pr(\tau_i|M, \theta, \gamma_i)$ where $\gamma_i$ assigns sequences to tips in $M$ is replaced by $\Pr(\gamma_i(d_i)|\tau_i, \alpha_i) \times \Pr(\tau_i |M,\theta)$ where $\gamma_i$ is now thought of as swapping rows in the data matrix. This is mathematically equivalent but does not work well in implementation.}
\end{itemize}

\subsection{The AlloppNET model}

The formula for the posterior probability for the AlloppNET model is similar to that for AlloppMUL, and is given by

\begin{eqnarray}\label{eq:AlloppNET}
\Pr(W, \theta, \tau, \alpha, \gamma | d) & \propto & \Pr(W|\lambda)\Pr(\lambda) \times \nonumber\\
                              &         & \Pr(\theta|\eta)\Pr(\eta) \times \nonumber\\
                              &         & \Pr(\gamma) \times \nonumber\\
                              &         & \prod_{i=1}^G \Pr(\tau_i|M_W, \theta, \gamma_i) \times \nonumber\\
                              &         & \prod_{i=1}^G \Pr(d_i|\tau_i, \alpha_i).
\end{eqnarray}
The network is denoted by $W$ and the multiply labelled tree derived from it is $M_W$. The other parameters are similar to those appearing in equation (\ref{eq:AlloppMUL}), but the meanings of $\lambda$ and $\theta$ are somewhat different. The terms for the permutation prior and the gene tree likelihood are as before, but the models differ in the meaning of the other terms, as described next.

\begin{itemize}
\item{The network prior is $\Pr(W|\lambda)\Pr(\lambda)$, and again, little is known about what an appropriate prior should be. The prior used here was designed using the `trees with legs' representation. Thus there is a diploid tree with two tips and unknown age, a tetraploid subtree with known age equal to the hybridization time, and the two legs. The priors for the diploid tree and the allotetraploid subtree both use a birth-death model \citep{Gernhard:2008} with the ratio of extinction rate to speciation rate fixed at 0.8, so that these priors can be regarded as one-parameter models. Furthermore, the diversification rate (speciation rate minus extinction rate) is assumed to be the same for both trees and is the single parameter $\lambda$ to be estimated. The hybridization time has a uniform prior on the interval between the diploid root and the present. Finally, the topologies for the legs were given probabilities of 1/3 for scenario $A$, 1/6 for $B$ and for its mirror image, 1/6 for $C$ and for its mirror image, and the two node times given uniform prior on the interval between hybridization time and the diploid root.}

\item{In the population prior $\Pr(\theta|\eta)\Pr(\eta)$, there is one value at each tip, one at the root-ward end of each edge in the network, and one just after the hybridization event. This allows the population to change discontinuously at hybridization. As in AlloppMUL, $\theta$ is the vector of these values and the prior for $\theta$ used here were similar to those typically used by *BEAST.}

\item{The formula for $\Pr(\tau_i|M_W, \theta, \gamma_i)$ is similar to that for AlloppMUL. The main difference is that the multiply labelled tree $M_W$ is derived from the structure of the underlying network $W$. Note also that since the population is allowed to change discontinuously at hybridization, the coalescent formula must be applied separately to the intervals before and after hybridization.}
\end{itemize}

\subsection{Implementation of the allopolyploid models}

The two models were implemented in BEAST. They both use the existing models for gene trees, and therefore can use all the models for substitution, site rate heterogeneity, and clock rates available in BEAST. They have MCMC operators which can swap the assignments of a pair of sequences for an individual, or various groupings of individuals. The assignment of one or two sequences to individuals, and of individuals to species is specified in the input XML file.

For AlloppNET, the species network is modeled as a set of trees with legs, which is then converted to a MUL-tree representation as needed. MCMC operators were designed to explore the space of species network topologies and node times, for example moving the legs, changing the hybridization time, and changing the tetraploid subtree. The MUL-tree representation is used to calculate the coalescent likelihood. Since the MUL-tree is always derived from the underlying network, it automatically has the equality constraints on the topology, node times and populations between the pair of tetraploid subtrees.

\section{Simulations and empirical data}

\subsection{Scenarios used in simulations}

Six scenarios, as shown in Figures 2 and 3 were used to simulate DNA sequences. Each scenario represents a `true' MUL-tree. Heights are in units of expected substitutions per site. Population sizes are effective numbers of gene copies within diploid populations (twice the number of individuals), or numbers of gene copies with the same diploid parent, for allotetraploid populations. If the effective population size is $S$, the probability of coalescence between a pair of gene copies is $1/S$ per generation. Population sizes are 100,000 at tips, and at root-ward ends of branches, and 200,000 at tip-ward ends of internal branches and at the root. All genes have length 500.

\begin{figure}[!hbtp]
\resizebox{1.00\hsize}{!}{\includegraphics*{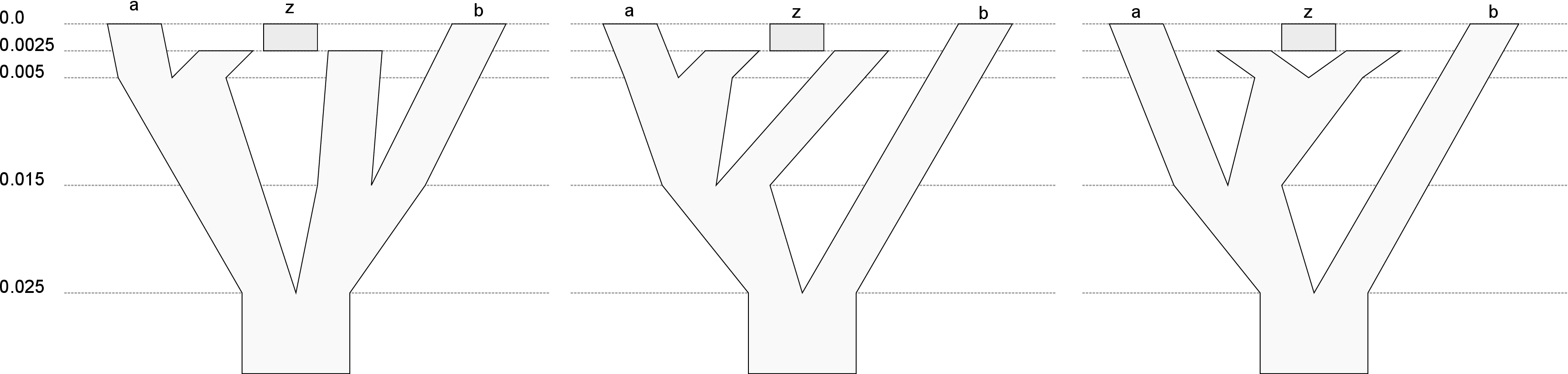}}
\caption{\textsf{Scenarios $A1$, $B1$, $C1$. Heights are in expected substitutions per site.}}
\end{figure}

\begin{figure}[!hbtp]
\resizebox{1.00\hsize}{!}{\includegraphics*{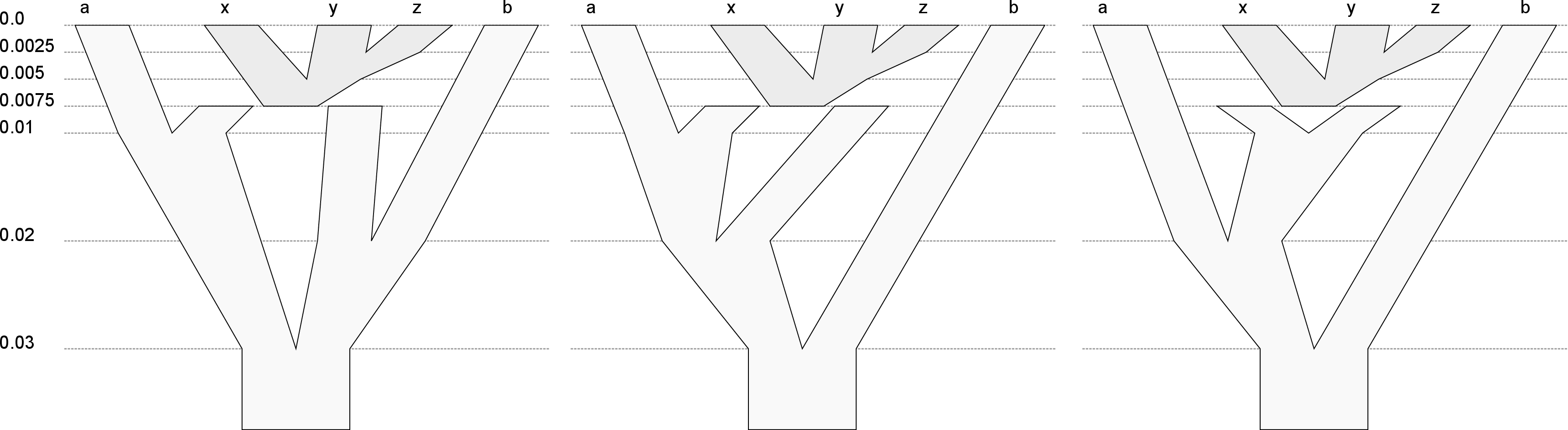}}
\caption{\textsf{Scenarios $A3$, $B3$, $C3$. Heights are in expected substitutions per site.}}
\end{figure}

These six scenarios were each tested with the number of genes $G$ equal to 1, 3 and 9, and the number of individuals $N$ per species equal to 1 and 3. The mutation rates $T$ were set to $4 \times 10^{-8}$ and $8 \times 10^{-8}$ for scenarios $A3$, $B3$, and $C3$ and $4 \times 10^{-8}$, $8 \times 10^{-8}$, and $1.6 \times 10^{-7}$ for scenarios $A1$, $B1$, and $C1$. The $T$ values are in expected substitutions per site per generation. Scenarios $A1$, $B1$, and $C1$ have a root height of 0.025. Scenarios $A3$, $B3$, and $C3$ have a root height of 0.03. Changing $T$ while keeping this height fixed changes the number of generations the tree represents. For example $T=4\times 10^{-8}$ in scenarios $A3$, $B3$, and $C3$ means 0.03/$4 \times 10^{-8}$ = 750,000 generations root to tip. In general, increasing $G$ and $N$ is expected to increase accuracy since there is more data, while increasing $T$ is expected to decrease accuracy since incomplete lineage sorting becomes more common.

\subsection{Implementation of simulations}

The simulations and the analyses of results were implemented in R \citep{R:08}. The input scenarios, as shown in Figures 2 and 3, were converted to a MUL-tree, then gene tree topologies and coalescent times were simulated according to the coalescent model within branches. The sequences were then generated using Seq-Gen \citep{seqgen}. 

Strict clock branch rates, and no site rate heterogeneity were assumed in both the simulations and the inference. In the simulations, equal clock rates for all genes were assumed, whereas in the inference, the clock rate for one gene was fixed to 1.0, and the others were estimated. The HKY substitution model was assumed in simulations and the inference. In the simulations, the substitution model parameters kappa were set to 3, and the frequencies set to .3 for A and T, and .2 for C and G (Seq-Gen was called with parameters \verb+-t3.0 -f0.3,0.2,0.2,0.3+). These were estimated in the inference. For the inference, the priors for the population parameter $\eta$ and the parameter $\lambda$ appearing in the network prior were the `OneOnX' distribution, which is improper, and the priors for relative clock rates were a very diffuse gamma distribution with mean 1 and shape parameter 0.1. These are the default priors used in *BEAST.

BEAST XML files were generated containing the simulated sequences for the AlloppNET and AlloppMUL models. There were 18 values for the triple ($G,N,T$) for scenarios $A1$, $B1$, and $C1$ and 12 for scenarios $A3$, $B3$, and $C3$, making $54+36=90$ configurations in total. For each of these, 20 replicates were simulated and run for 1.5 million generations in BEAST using both models, making a total of $90 \times 2 \times 20 = 3600$ BEAST runs. MUL-trees were sampled every 1000 generations, and the first 501 samples (of 1501) discarded as burn-in. For all six scenarios, three values for $G$, and two for $N$ were used.

\subsection{Empirical data}

The empirical data was analysed using very similar assumptions to the simulated data. Strict clock branch rates, and no site rate heterogeneity were assumed. The clock rate for one gene was fixed to 1.0, and the others were estimated. The HKY substitution model was used. For the \textit{Silene} data, the priors for $\eta$ and $\lambda$, and the priors for relative clock rates were the default priors used in *BEAST. For the \textit{Pachycladon} data, these priors were changed, as described later.

Two sets of empirical data were analyzed. The first comes from a study \citep{Joly:2009} of the genus \textit{Pachycladon} (Brassicaceae) which consists of eight species, plus a number of diploids. This study showed that the \textit{Pachycladon} genus originated from an allopolyploidization which is estimated to have occurred between 1.6 and 0.8 Mya. For the present analysis, the eight \textit{Pachycladon} species together with the two diploids \textit{Arabidopsis thaliana} and \textit{Lepidium apelatum} were used. There is one individual for each species, and five genes. There was a substantial amount of missing data: out of a possible 90 sequences (assuming that every diploid genome contributes with one allele each), 41 were unavailable.

The second data set comes from \textit{Silene} (Caryophyllaceae). There is one allotetraploid \textit{S.involucrata}, here labeled `Si'. The species delimitations for the diploids are currently under investigation (Petri and Oxelman, unpublished). For this analysis, the taxa \textit{S.ajanensis}, \textit{S.linnaeana}, \textit{S.samojedora}, and  \textit{S.villosula} were grouped together, and labeled `Salsv', and \textit{S.uralensis} and \textit{S.violascens} were merged and labeled `Suw'. The sequence data comes from the four low-copy nuclear genes NRPA2, NRPB2, NRPD2a, and NRPD2b \citep[Petri and Oxelman, unpublished]{Popp:2005}. There were 12 individuals from Salsv, 8 from Suw, and 4 from Si. Out of a possible 112 sequences, 34 were missing.

\section{Results}

\subsection{Simulations}

Tables \ref{tab:ResultsA1}-\ref{tab:ResultsC3} show the results as number of times the correct topology of the MUL-tree was recovered as the concensus tree. For scenarios $A1$ and $A3$, the legs are regarded as matching the true topology if one leg joins to diploid $a$ and the other to diploid $b$. In scenarios $B1$ and $B3$, the legs are regarded as matching the true topology if both legs join diploid $a$ (that is, the order of the two joins is not considered). In scenarios $C1$ and $C3$, the legs are  regarded as matching the true topology as long as the two legs join one another and their common ancestor joins diploid $a$; in this case the legs are indistinguishable from one another. For scenarios $A3$, $B3$, and $C3$, the `relaxed' match means that the legs match the true topology in this manner. The `strict match' uses the same criterion for the legs but requires that the topology of the tetraploid subtree also matches the true topology.

\begin{table}[!p]
\begin{tabular}{| c   c   c   c  c  c  c |}
\hline
         & \multicolumn{2}{ c }{625,000}  & \multicolumn{2}{c}{312,500} & \multicolumn{2}{c|}{156,250}  \\
$G$, $N$ & MUL & NET  & MUL & NET & MUL & NET  \\
\hline
1,1      & 14  & 14 & 9 & 11 & 5 & 9\\
1,3      & 14  & 15 & 13 & 13 & 12 & 12\\
3,1      & 18  & 17 & 15 & 13 & 14 & 10\\
3,3      & 19  & 20 & 17 & 16 & 18 & 19\\
9,1      & 20  & 20 & 19 & 18 & 15 & 14\\
9,3      & 20  & 20 & 20 & 20 & 20 & 20\\
\hline
\end{tabular}
\caption
{{\small Results for simulated scenario $A1$. The number of generations root to tip is shown in the first row. The first column shows the number of genes $G$ and the number of individuals per species $N$. MUL and NET refer to the two models AlloppMUL and AlloppNET. The figures in the main part of the table are the number of correct topologies out of 20 replicates.}\label{tab:ResultsA1}}
\end{table}

\begin{table}[!p]
\begin{tabular}{| c   c   c   c  c  c  c |}
\hline
     & \multicolumn{2}{ c }{625,000}  & \multicolumn{2}{ c }{312,500} & \multicolumn{2}{ c|}{156,250}  \\
$G$, $N$ & MUL & NET  & MUL & NET & MUL & NET  \\
\hline
1,1  & 15 & 18 & 11 & 15 & 6 & 16 \\
1,3  & 16 & 18 & 13 & 17 & 13 & 14\\
3,1  & 17 & 19 & 15 & 18 & 13 & 16 \\
3,3  & 16 & 17 & 18 & 18 & 18 & 19\\
9,1  & 19 & 20 & 19 & 20 & 19 & 18\\
9,3  & 20 & 20 & 20 & 20 & 20 & 20\\
\hline
\end{tabular}
\caption{{\small Results for simulated scenario $B1$. Details as Table \ref{tab:ResultsA1}.}\label{tab:ResultsB1}}
\end{table}

\begin{table}[!p]
\begin{tabular}{| c   c   c   c  c  c  c |}
\hline
     & \multicolumn{2}{ c }{625,000}  & \multicolumn{2}{ c }{312,500} & \multicolumn{2}{ c|}{156,250}  \\
G, N & MUL & NET  & MUL & NET & MUL & NET  \\
\hline
1,1  & 11 & 15 & 6 & 9 & 3 & 8\\
1,3  & 16 & 19 & 11 & 14 & 15 & 16\\
3,1  & 19 & 20 & 15 & 19 & 13 & 14\\
3,3  & 18 & 19 & 19 & 19 & 19 & 19\\
9,1  & 20 & 20 & 19 & 20 & 20 & 20\\
9,3  & 20 & 20 & 20 & 20 & 20 & 20\\
\hline
\end{tabular}
\caption{{\small Results for simulated scenario $C1$. Details as Table \ref{tab:ResultsA1}.}\label{tab:ResultsC1}}
\end{table}

\begin{table}[!p]
\begin{tabular}{| c   c   c   c  c   c   c   c  c |}
\hline
     &  \multicolumn{4}{ c }{Strict match} & \multicolumn{4}{ c|}{Relaxed match }\\
     & \multicolumn{2}{ c }{750,000}  & \multicolumn{2}{ c }{375,000} & \multicolumn{2}{ c }{750,000}  & \multicolumn{2}{ c|}{375,000}  \\
G, N & MUL & NET  & MUL & NET  & MUL & NET  & MUL & NET  \\
\hline
1,1  & 3 & 3    & 0 & 1   & 4  & 6    & 1  & 3  \\
1,3  & 8 & 13   & 8 & 13  & 12 & 14   & 11 & 16  \\
3,1  & 11 & 17  & 5 & 8   & 18 & 19   & 12 & 14   \\
3,3  & 18 & 19  & 12 & 18 & 20 & 20   & 18 & 19  \\
9,1  & 14 & 19  & 6 & 13  & 20 & 20   & 19 & 20 \\
9,3  & 19 & 20  & 17 & 20 & 20 & 20   & 18 & 20  \\
\hline
\end{tabular}
\caption{{\small Results for simulated scenario $A3$. The number of generations root to tip is shown in the second row. The first column shows the number of genes $G$ and the number of individuals per species $N$. MUL and NET refer to the two models AlloppMUL and AlloppNET. The figures in the main part of the table are the number of topologies out of 20 replicates which match the correct topology: columns 2-5 for strict match, and 6-9 for relaxed match; see text for details.}\label{tab:ResultsA3}}
\end{table}

\begin{table}[!p]
\begin{tabular}{| c   c   c   c  c   c   c   c  c |}
\hline
     &  \multicolumn{4}{ c }{Strict match} & \multicolumn{4}{ c|}{Relaxed match }\\
     & \multicolumn{2}{ c }{750,000}  & \multicolumn{2}{ c }{375,000} & \multicolumn{2}{ c }{750,000}  & \multicolumn{2}{ c|}{375,000}  \\
G, N & MUL & NET  & MUL & NET  & MUL & NET  & MUL & NET  \\
\hline
1,1  & 2 & 11  & 1  & 10  & 3  & 17 & 1  & 18 \\
1,3  & 8 & 16  & 11 & 14  & 10 & 18 & 15 & 18\\
3,1  & 13 & 17 & 7  & 12  & 20 & 19 & 14 & 18   \\
3,3  & 15 & 19 & 17 & 19  & 20 & 19 & 20 & 20 \\
9,1  & 15 & 18 & 10 & 14  & 20 & 20 & 20 & 20\\
9,3  & 19 & 20 & 18 & 20  & 20 & 20 & 20 & 20 \\
\hline
\end{tabular}
\caption{{\small Results for simulated scenario $B3$. Details as Table \ref{tab:ResultsA3}.}\label{tab:ResultsB3}}
\end{table}

\begin{table}[!p]
\begin{tabular}{| c   c   c   c  c   c   c   c  c |}
\hline
     &  \multicolumn{4}{ c }{Strict match} & \multicolumn{4}{ c|}{Relaxed match }\\
     & \multicolumn{2}{ c }{750,000}  & \multicolumn{2}{ c }{375,000} & \multicolumn{2}{ c }{750,000}  & \multicolumn{2}{ c|}{375,000}  \\
G, N & MUL & NET  & MUL & NET  & MUL & NET  & MUL & NET  \\
\hline
1,1  & 3  & 11 & 2  &  7 & 6 & 20   & 5 & 16 \\
1,3  & 5  & 15 & 7  & 17 & 6 & 17   & 8 & 17 \\
3,1  & 13 & 16 & 4  & 14 & 17 & 20   & 10 & 20   \\
3,3  & 18 & 20 & 11 & 19 & 18 & 20   & 17 & 20 \\
9,1  & 16 & 18 & 13 & 18 & 19 & 20   & 17 & 20 \\
9,3  & 20 & 20 & 18 & 20 & 20 & 20   & 20 & 20 \\
\hline
\end{tabular}
\caption{{\small Results for simulated scenario $C3$. Details as Table \ref{tab:ResultsA3}.}\label{tab:ResultsC3}}
\end{table}


The values of $T$ has a major impact on the difficulty of the problem, as expected. $T$ determines the lengths of the branches units of generations. The key quantity is the ratio of the length of the branch measured in generations to the population size. Increasing $G$ and $N$ improves the accuracy, also as expected. In general, it appears that increasing $N$ is most useful when the branches that need to be resolved are recent, whereas increasing $G$ is most useful when the branches that need to be resolved are more ancient. This was also observed in similar scenarios with the same topology but different node times (results not shown). If the important branches are deep in the tree, the sequences from different individuals usually coalesce too soon (going back in time) to be useful.

For $A1$, $B1$, and $C1$ together, AlloppNET is better in 28 cases, AlloppMUL is better in 7 cases, with 19 draws. Thus AlloppNET seems slightly better on the scenarios with one tetraploid but the difference is of little practical importance. For the scenarios with three tetraploids AlloppNET is clearly a lot better: AlloppMUL appears to need around twice the amount of data to achieve similar accuracy. AlloppNET took around 1.5 times as much computational time per generation as AlloppMUL.

The sampled values of the population size parameters have high variance and are highly skewed, and so it seems preferable to work with the logarithms of these values. Estimates of the logarithms of population size parameters for scenario $B1$ are shown in Figure 4. Note that in the prior the mean of the tip population size parameters is twice that of those elsewhere in the tree, while in the simulations, all the parameters are equal. In most of the replicates, the influence of the prior was clear, even with $G=9$ and $N=3$, with the tip populations overestimated and the rest underestimated. This indicates that there is little information in the data about the population sizes in individual branches. 

\begin{figure}[!hbtp]
\resizebox{1.00\hsize}{!}{\includegraphics*{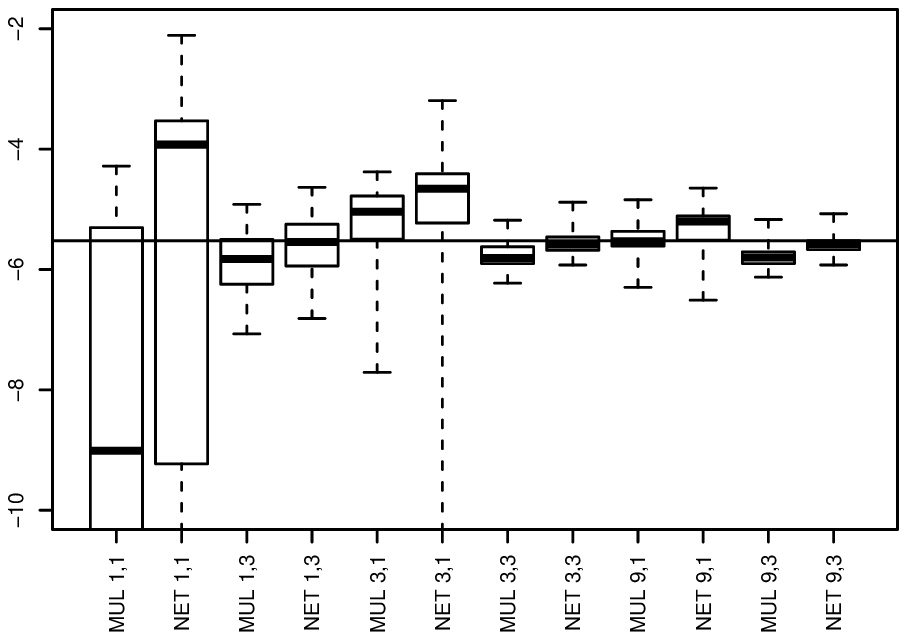}}
\caption{\textsf{Estimates (in log-space) of population parameters for scenario $B1$, with $T=8 \times 10^{-8}$. The values used for each boxplot are the means, averaged over branches and over samples in the MCMC chain, of the logarithms of the population parameters. Each boxplot shows the results from 20 replicates for a model and particular numbers of genes and individuals. For example `NET 3,1' means the AlloppNET model with 3 genes and one individual per species. The boxes show interquartile ranges, and the whiskers show the extremes of the ranges. The horizontal line is at the true value of $\log(0.008)$}.}
\end{figure}

Estimates of the root heights of the MUL-tree for scenario $B1$ are shown in Figure 5. Note that in three cases where there is little data, the estimates from AlloppMUL are occasionally extremely small. This may be due to a failure of the MCMC process to converge. 

\begin{figure}[!hbtp]
\resizebox{1.00\hsize}{!}{\includegraphics*{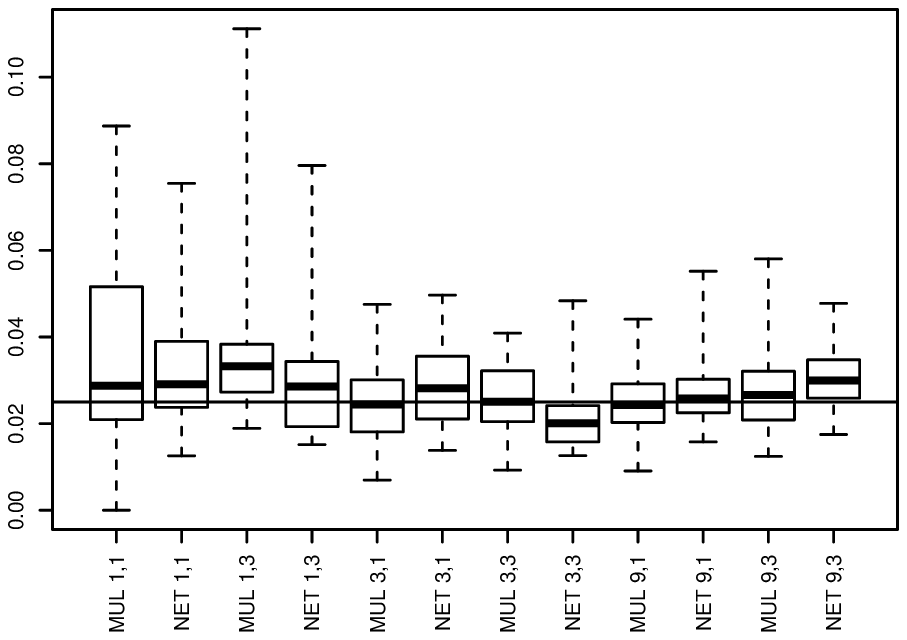}}
\caption{\textsf{Estimates of root heights for scenario $B1$, with $T=8 \times 10^{-8}$. The horizontal line is at the true value, $0.025$. Other details as Figure 4.}}
\end{figure}

\subsection{Empirical data}

\subsubsection{Pachycladon data} There were some convergence problems with the AlloppMUL model when using the default priors for the parameters $\eta$, $\lambda$, and the relative clock rates. Convergence often failed to occur after 10 million generations, and the results were dubious even after 100 million generations. Changing the priors to more realistic ones appeared to improve this behavior considerably. Log-normal distributions were used, and their parameters, in log-space, were as follows: for $\eta$, a mean of -6.0 and a standard deviation of 1.5; for $\lambda$, a mean of 4.6 and a standard deviation of 2.0; and for the relative clock rates a mean of 0.0 and a standard deviation of 1.0. No convergence problems where observed when using the AlloppNET model, and convergence to the expected topology usually appeared to occur within two million generations. It is not surprising that the AlloppMUL model has more difficulty than AlloppNET with this data set, since with eight allotetraploids, there are far more possible topologies allowed by the model, as well as more node times and population parameters to estimate.

The results reported here use the log-normal priors and a run length of 100 million generations, with the first half discarded as burn-in. They are shown in Figures 6 and 7. Both models infer that the \textit{Pachycladon} genomes are both closer to \textit{Arabidopsis thaliana} than \textit{Lepidium apelatum} and that the they join the \textit{Arabidopsis} lineage separately. This is the result expected from the previous analysis by \cite{Joly:2009}, and corresponds to scenario $B$ in the simulations. 

\begin{figure}[!hbtp]
\resizebox{1.00\hsize}{!}{\includegraphics*{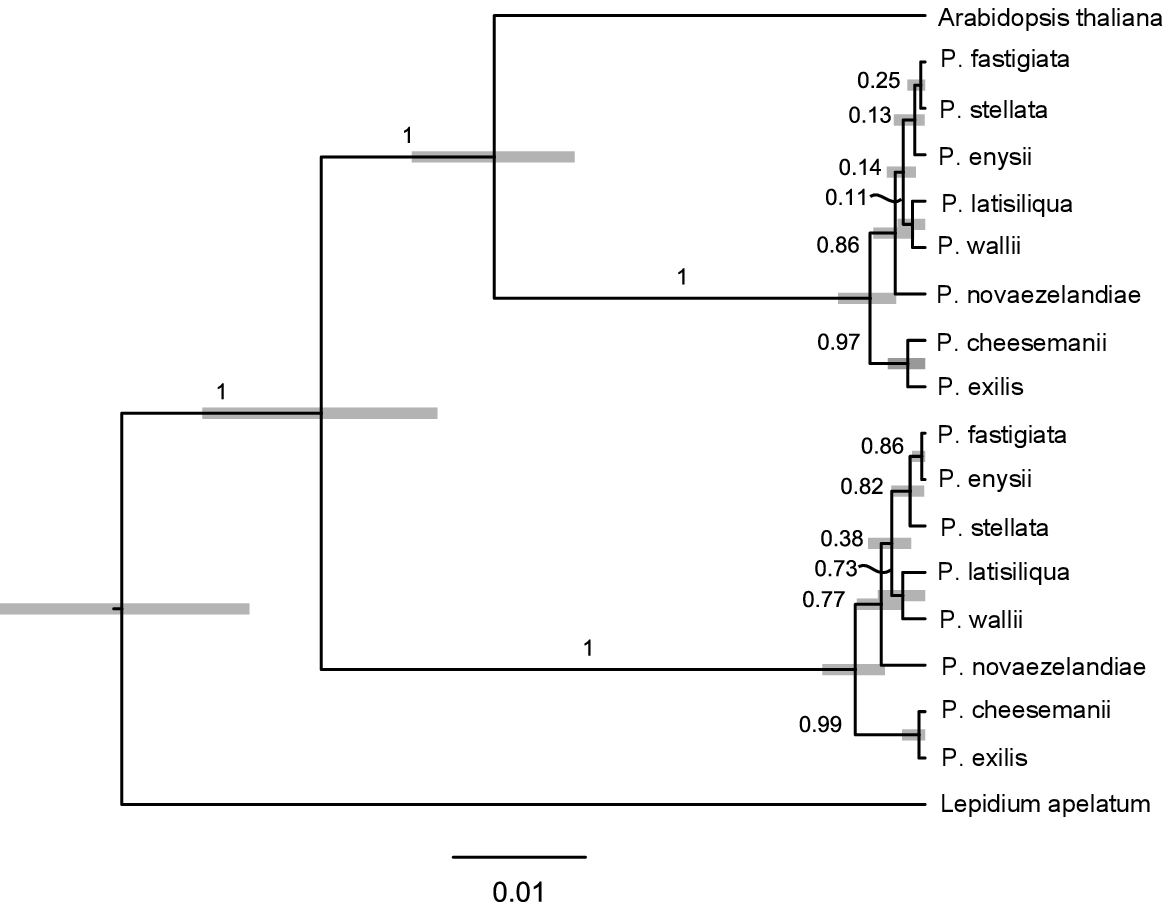}}
\caption{\textsf{Pachycladon MUL-tree estimated using AlloppMUL. Heights are in expected substitutions per site based on the CHS gene. The grey bars at nodes indicate 95\% HPD intervals for node height. Posterior clade probabilities are shown as numbers for internal branches}.}
\end{figure}

\begin{figure}[!hbtp]
\resizebox{1.00\hsize}{!}{\includegraphics*{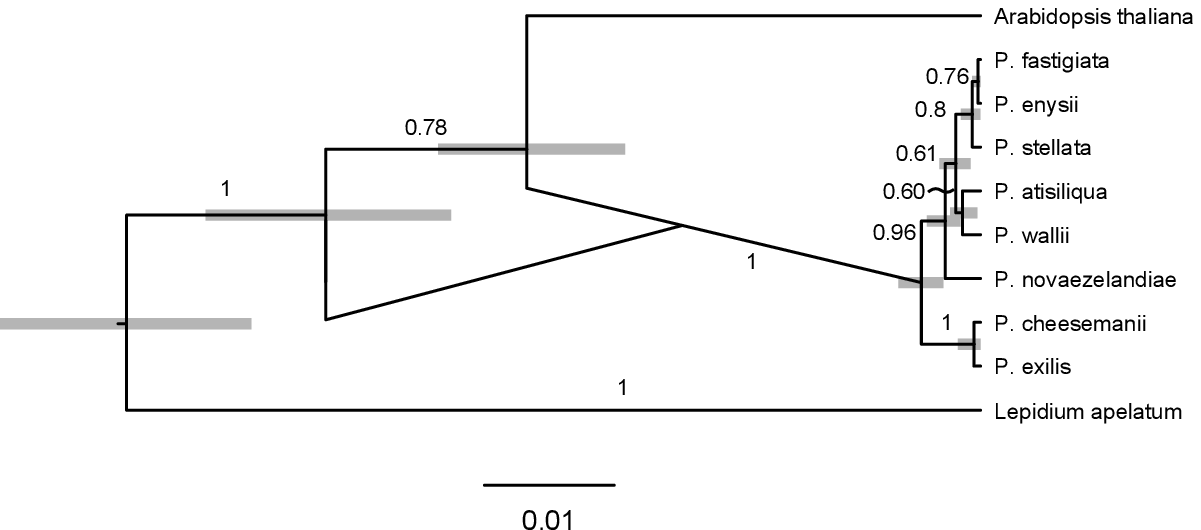}}
\caption{\textsf{Pachycladon network estimated using AlloppNET. Heights are in expected substitutions per site based on the CHS gene. The position of the hybridization node does not represent an estimate of the hybridization time. The grey bars at nodes indicate 95\% HPD intervals for node height. Posterior clade probabilities are shown as numbers for internal branches.}}
\end{figure}

There is no strong disagreement between the topology of the \textit{Pachycladon} subtree estimated using AlloppMUL, AlloppNET, or the CHS gene tree from \cite{Joly:2009}. However, the AlloppNET subtree is fully resolved, whereas the CHS tree is not. The CHS gene is the only gene that was sequenced for all eight species. Although the other genes appear to contribute with little topological information, AlloppNET is capable of taking this information into account.


The mean $\zeta$ of the logarithms of the population parameters along branches were calculated for each MCMC sample. For AlloppMUL the 95\% HPD interval for $\zeta$ was (-8.5, -6.2) with a median and mean of -7.3. For AlloppNET the 95\% HPD interval for $\zeta$ was (-8.8, -6.2) with a median and mean of -7.5. A very approximate calculation can be made for the average population. Taking $\exp(-7.4) \simeq 6 \times 10^{-4}$ as a typical value for a population parameter along a branch, and an estimate of the mutation rate as approximately $3 \times 10^{-8}$ per site per generation from \cite{Joly:2009}, the number of gene copies in a typical population is approximately 20000, and so the number of individuals is estimated at 10000. The 95\% HPD interval for the number of individuals is approximately (2500,40000).

\subsubsection{Silene data} No convergence problems were observed here and the performance of the two models was very similar. The two trees are shown in Figures 8 and 9. For AlloppMUL the 95\% HPD interval for $\zeta$ was (-7.3, -6.2) with a median and mean of -6.7. For AlloppNET the 95\% HPD interval for $\zeta$ was (-7.0, -5.7) with a median and mean of -6.3. 

\begin{figure}[!hbtp]
\resizebox{0.5\hsize}{!}{\includegraphics*{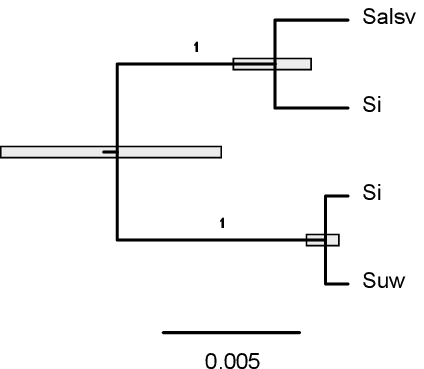}}
\caption{\textsf{Silene MUL-tree estimated using AlloppMUL. Heights are in expected substitutions per site based on the RPA2 gene.}}
\end{figure}

\begin{figure}[!hbtp]
\resizebox{0.5\hsize}{!}{\includegraphics*{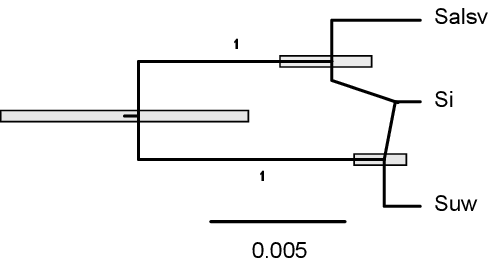}}
\caption{\textsf{Silene network estimated using AlloppNET. Heights are in expected substitutions per site based on the RPA2 gene. The position of the hybridization node does not represent an estimate of the hybridization time.}}
\end{figure}

\section{Discussion}

This paper represents a first step towards the statistical inference of allopolyploid networks. The two models are complementary in that AlloppMUL is applicable to more data sets, while AlloppNET is more powerful if one can restrict to two diploids and a single hybridization. Under the assumption that the constituent genomes of  an allopolyploid does not recombine, AlloppNET will reconstruct species trees under the multispecies coalescent, even if there is substantial amounts of missing data, as can be seen from the \textit{Pachycladon} example. AlloppMUL could be useful for situations where the number of hybridizations are unknown. The number of hybrdization could be inferred using methods like PADRE \citep{HuberOxelman:2006, LottEtal:2009, Marcussen:2012}. Both models are available as part of BEAST 1.7 \citep{BEAST17}, and can take advantage of the numerous models for sequence evolution within gene trees. There is currently no support for the models in Beauti, but R scripts are available to aid the construction of suitable XML files.

The AlloppNET model could be extended to deal with arbitrary numbers of diploids, and then to deal with an unknown number of hybridization events, both for allotetraploids and for higher ploidy levels. Designing a suitable prior for such networks is another open problem. Even from a purely mathematical view, with no concern for biological realism, it appears to be difficult to write down a density for all possible networks that might have given rise to a particular number of diploids and tetraploids, since the number of nodes in the network (and therefore the number of parameters) changes with the number of hybridizations.

\centerline{\textsc{Funding}}

This work was supported by the Swedish Research Council (621-2010-5623 to SS and 2009-5202 to BO) and the Centre for Theoretical Biology at the University of Gothenburg (GJ).

\centerline{\textsc{Acknowledgement}}

We thank Anna Petri for sharing some unpublished \textit{Silene} sequences.

\bibliographystyle{sysbio}
\bibliography{references}

\end{document}